# Valence electron distribution in MgB$_2$ studied by accurate diffraction measurements and first principle calculations

Lijun Wu, Yimei Zhu, T. Vogt, Haibin Su, and J. W. Davenport

Brookhaven National Laboratory, Upton, New York 11973

J. Tafto

University of Oslo. P.O. Box 1048, Blindern, 0316 Oslo, Norway

## Abstract

We use synchrotron x-ray and precision electron diffraction techniques to determine accurately the structure factors of selected reflections that are sensitive to the valence electron distribution in the superconductor MgB$_2$. These values deviate significantly from calculated structure factors using the scattering factors of free atoms, but agree well with our calculated structure factors based on density functional theory. Having experimentally established the reliability of our first-principle calculated structure factors, we present electron density maps of the redistribution of the valence electrons that takes place when hypothetical free atoms of Mg and B in MgB$_2$ interact to form the real crystal.

The electronic structure of $MgB_2$ has received much attention since the discovery of superconductivity at the astonishing high temperature of 39 K [1]. This material is bound to be a model system in efforts to understand superconducting properties from microscopic first principle calculations. This statement is based on the fact that this remarkable material is structurally simple with only low Z elements in a small crystal unit cell with high symmetry, P6/mmm. It is important to recognize that Mg (Z=12) and B (Z=5) are light elements and therefore charge transfers lead to large changes in the total charge density on bonding. In x-ray crystallography, most investigations are based on overlapped spherical atomic charge densities, thus neglecting subtle bonding effects [2,3]. The structural simplicity of $MgB_2$ leads to the situation that just a few months after the discovery of its superconductivity, many first principles electronic structure calculations based on Density Functional Theory, DFT, appeared in the literature [4-7]. Virtually all the first-principles calculations that have been published show the same electronic structure and density of states, and thus also the same spatial distribution of electrons. Experimental techniques to verify the calculated bulk electronic structure calculations are x-ray absorption spectroscopy and electron energy loss spectroscopy, focusing on the fine structure near the absorption edge, and quantitative diffraction using x-rays or fast electrons. Spectroscopy and diffraction are complementary experimental tools. Diffraction techniques are experimentally more demanding than spectroscopic techniques since extremely accurate diffraction measurements are needed to move beyond determining the positions of the atoms (nuclei) within the crystal unit cell towards addressing the rearrangement of valence electrons caused by bonding between the atoms of the crystal. The interpretation of the spectroscopic data, on the other hand, poses an additional theoretical challenge because the ejection of core electrons forces the system into an excited state while DFT applies strictly only to the ground state. Several studies have recently addressed the electronic structure of $MgB_2$ using spectroscopic techniques [8-10].

In this work, we use a novel quantitative electron diffraction technique we have developed recently [11] to accurately measure the structure factors of low-order reflections that are very sensitive to the charge distribution in materials. We also use synchrotron x-ray diffraction for high-order reflections. DFT calculations of the electronic structure of $MgB_2$ are then tested by these diffraction measurements in order to study the rearrangement of the electron density as compared to the electron density in a hypothetical crystal with atoms having the electron distribution of free atoms (overlapped spherical atom densities, here termed as procrystal). Using x-ray diffraction we extract structure factors that are the Fourier components of the electron density in the crystal. With electron diffraction we determine the Fourier components of the electrostatic potential in the crystal. The final step of converting to x-ray structure factors is straightforward for short **g**-vectors, while for larger **g**-vectors the thermal parameters need to be determined with high accuracy. Likewise the electron density that we calculate by DFT can be represented by its Fourier components.

X-ray and electron diffraction are complementary. With x-rays we collect powder diffraction data of many reflections, while with electrons we can determine accurately selected structure factors, and due to its small probe, a tiny powdered grain or a crystal smaller than 100 nm can be easily studied as a single crystal. Electron diffraction is particularly sensitive to valence electron distribution in crystals at short reciprocal vectors, and thus the structure factors determined from electron diffraction at small scattering angles give valuable information about electron transfer over longer distances in the crystal unit cell. Measuring accurately just a few of these structure factors provides a useful test of electronic structure calculations of the electron density. Electron density maps of $MgB_2$ have been published based on electronic structure calculations and on synchrotron x-ray diffraction on powder [12] and single crystal measurements [13], but the single crystals grown so far are reported to be off stoichiometry[13]. Neither measurement of electron diffraction nor quantitative comparisons have been made between experiments and the electron density obtained from the first principles calculations. The objective of this Letter is to report such measurements and comparisons that are best done in reciprocal space relying on structure factors.

The diffraction experiments were done at room temperature on a well-characterized powder of $MgB_2$ [14]. In the electron diffraction experiments, we focused the electron probe within the individual micrometer sized crystal grains. We operated our 3000F JEOL transmission electron microscope that is equipped with a field emission gun and a Gatan energy filter, at the accelerating voltage of 300 keV which corresponds to an electron wavelength of 1.97 pm. We use convergent beam electron diffraction to accurately determine the electron structure factors of selected reflections. The conventional way of recording a convergent beam electron diffraction pattern is to focus the fast electrons on the specimen [15], resulting in recording the diffraction pattern at a constant thickness, but over the range of incident beam directions defined by the convergent beam angle. We refer to this procedure as Conventional Convergent Beam Electron Diffraction (CCBED). An energy-filtered CCBED pattern with the 001 reflection at the Bragg position is shown in fig. 1a. A more robust experimental procedure is to focus the electron probe of diameter down to 1 nm some 100μm above the specimen [11]. Each convergent beam disk will now be a shadow image of the specimen, and we achieve PArallel Recording Of Dark-field Images, PARODI. An area of diameter of the order 1μm, depending on the convergent beam angle and the distance from the crossover to the specimen, can be illuminated with electrons. If the specimen varies in thickness over the illuminated area, we thus record the diffraction intensities as a function of thickness in addition to the incident beam direction. When compared to CCBED, PARODI adds a new dimension, the thickness, to the experiment as shown in fig.1d thus giving additional information in each diffraction disk which is extremely sensitive to small changes of the crystal and electronic structure under investigation. In this study we excluded the inelastically scattered electrons from the diffraction patterns using an energy filter with an energy window of 5 eV.

To extract structure factors from CCBED as well as PARODI patterns, we perform dynamical Bloch wave calculations of the diffraction intensities as a function of thickness and crystallographic direction of the incident beam. For $MgB_2$ we know the crystal structure and start by using structure factors of the procrystal, i.e. based on the scattering factors of free atoms. These scattering factors are available in the literature based on Dirac-Fock calculations [16]. Then iteratively we change the structure factors until we arrive at the best agreement with experiment. In these calculations we used the room temperature lattice parameters determined by Vogt et al.[14] with a= 0.3086nm and c= 0.3521 nm and the Debye-Waller factor B= 0.4 for both boron and magnesium [17]. Fig. 1b-c and e-f show calculated CCBED and PARODI patterns, assuming the scattering amplitudes of free atoms, and using the magnitudes of the structure factors giving the best fit to the experiments in fig. 1a and 1d. Using CCBED as well as PARODI resulted in an electron structure factor $F_{001}$= -0.20Å (fig.1 b) as compared to –0.91Å (fig.1 c) for the procrystal. We did similar experiments and fitting for the 100, 101 and 002 reflections. The measured, and the calculated structure factors based on the procrystal for the four reflections are shown in table 1 together with the x-ray structure factors that we arrive at from the electron diffraction experiments after having converted the electron structure factors using the following relationship between electron and x-ray structure factors

$$F_g^x = [\sum_i Z_i exp(-B\frac{g^2}{4}) exp(-2\pi i g \cdot r)] - (C\frac{g^2}{4} F_g)$$

Here B is the Debye-Waller factor, $C = 8\pi \varepsilon_0 h^2 / m_e e^2 = 41.78$ when the electron structure factor, $F_g$, and the reciprocal lattice vector, g, are given in Angstrom and reciprocal Angstrom units, respectively.

In table 1, we notice, relative to the procrystal, the larger change in the 001 and 100 electron structure factors than in the converted x-ray structure factors for these low order reflections. This is typical for reflections at small scattering angles. The reason is that the incident fast electrons interact with the electrostatic potential that includes the contribution from the positive charge at the nucleus. This results in large changes in the electron scattering amplitudes of the atoms at low scattering angles when valence electrons are redistributed. Thus electron structure factors for short reciprocal vectors are highly sensitive to charge transfer. We have previously shown that this is the case for the high temperature oxide superconductors [11] that have large c-axes and thus some reciprocal vectors that are much shorter than the shortest found in $MgB_2$.

First principles calculations of the electron density and structure factors were carried out using DFT. The DFT equations were solved using the Full Potential Linear Augmented Plane Wave method, FLAPW, as embodied in the WIEN code [18]. For the exchange correlation potential we used the Generalized Gradient Approximation of Perdew [19]. The lattice parameters corresponding to a minimum in the energy were a=0.3081 nm and c=0.3528 nm in good agreement with experiment and other DFT calculations. The band structure and density of states were also in good agreement with previous work [4-7]. We note that the WIEN code is a full potential implementation of the density functional equations - so for example the charge density is not constrained to be the overlap of spherical charge densities but rather contains whatever non spherical contributions are allowed by the crystal symmetry. Further, the charge density is determined self consistently and bears no particular relation to overlapped atomic densities.

To study reflections further out in reciprocal space we did powder x-ray diffraction measurements on beam line X7A at the National Synchrotron Light Source (NSLS) at BNL using x-rays of wavelength 79.99108 nm, fig.2a. The sample was contained in a 0.2 mm capillary which was spinning during the measurement at a speed of about 1 Hz to eliminate preferred orientation. The data were refined using the Rietveld analysis program PROFIL. In the initial refinement using the scattering factors of free atoms, a rather unsatisfactory fit was achieved as seen in the difference pattern in fig 2b. Subsequent refinement using the scattering factor of Mg2+ made a significant improvement, and the refinement was further improved by introducing additional electrons into the boron layer by partially replacing the scattering factor of the free boron atom with a carbon scattering factor, fig 2c [20], suggesting a considerable amount of charge transfer from Mg to B atoms. We note that in x-ray refinements too much electron density in the B-layer can be compensated for by introducing a Mg deficiency, thus creating an artifact due to using the wrong scattering factors and not necessarily an indication of non-stoichiometry.

In table 2 we compare experimental and calculated structure factors after having adjusted to T=0K for the experimental values by using the Debye-Waller factor of 0.4 for both Mg and B. Note that the experimental structure factors are much closer to the self consistent DFT results than those calculated assuming free atoms. We see, by comparing structure factors, also consistent with the x-ray refinement, that there is charge transfer away from the Mg plane towards the B plane in that there is a significant reduction of the 001 experimental structure factor relative to the one for the procrystal. The 002 structure factor is virtually the same for the electron diffraction experiments and for the procrystal, suggestive of the charge has moved the whole way from the Mg plane to the B plane.

In order to make quantitative comparison with experiments and with other calculations we first synthesized the difference density from the calculated DFT structure factors and those of the procrystal:

$$\Delta\rho(\vec{r}) = \rho(\vec{r})_{DFT} - \rho(\vec{r})_{PRO} = \frac{1}{V}\sum_{g}(F_{g_{DFT}} - F_{g_{PRO}})\exp(-2\pi i \vec{g}\cdot\vec{r}),$$

where the sum was carried out for $|\vec{g}| \leq (2\text{Å})^{-1}$, which corresponds to nearly 1000 reflections when the multiplicity of the crystal symmetry is considered. In fig.3 we present difference electron density maps through different planes in real space. Fig. 3a and fig 3b show cuts normal to the c-axis through the B-planes and Mg planes, respectively. Fig 3c shows the difference map through the atoms in the 110 plane, and fig. 3d a similar map after replacing the 001, 100, 101 and 002 structure factors in the DFT calculations with those determined by electron diffraction. In this work we want to provide robust tests for the DFT calculations by keeping the number of adjustable parameters to a minimum. Apart from the structure factors that we are aiming at, the only adjustable parameter using CCBED is the crystal thickness at the area we focus the electron probe, and in PARODI the angle of the crystal wedge. The structure factors that are most reliably tested experimentally are the innermost that are negligibly influenced by thermal parameters, and these are the structure factors that we can determine with high accuracy using quantitative electron diffraction. This is seen in table 1 where we note that the experimental electron structure factor of the 001 reflection is only 20 % of that for the procrystal, i.e. the hypothetical crystal with no rearrangement of the valence electrons. The similar percentage for the 100 reflection is 70%. However, for the majority of the reflections further out in reciprocal space, these deviations typically range between 98 and 102 % as they do for the x-ray structure factors based on comparison between DFT and the procrystal calculations. With the electron diffraction techniques we use here we measure the structure factors on an absolute scale, and are able to assess the accuracy of the fit by doing dynamical calculations with different values of the structure factors.

The four structure factors we have determined experimentally are in reasonable agreement with first principles DFT calculations, and the 001 structure factor that we have measured precisely using electron diffraction is in excellent agreement with these calculations suggesting that the DFT calculations we performed are adequate in addressing the redistribution of the valence electrons.

It is far from trivial to assign ionic charges to atoms from charge density maps. The major reason is the overlap between electrons that are attributed to the different ions in a crystal. A way of assigning the charge is to look at the magnitudes of the structure factors. The observation using electron diffraction that the 001 structure factor is 0.60 less than that calculated for the procrystal, table 2, is consistent with 2 electrons from each Mg atom having moved to the B plane: The x-ray scattering factor of $Mg^{2+}$ is 0.30 less than of a free Mg atom [15], and when these electrons move to the boron plane they scatter $\pi$ out of phase with the Mg ions, resulting in a total reduction of the 001 structure factor of 0.60 relative to the procrystal. On this crucial point our study does not agree with the previous study using the maximum entropy method [12], reporting that only half of the electrons that leave the Mg atoms move to the B-layer at 15K and none of them at room temperature [11].

In fig 4 we present a three-dimensional electron density map of the redistribution of the valence electrons in $MgB_2$, defined as the self consistent density minus that from overlapping spherical atomic densities based on the DFT calculations with structure factors up to $sin\theta/\lambda= 0.25$. A color scheme of blue-yellow-purple was used to represent charge depletion and excess of charge. The characteristic build up of bond charge between the B atoms is clearly apparent and this charge comes from the Mg as well as the B atoms.

As indicated by our experimentally tested DFT calculations, electrons are drained away from the Mg atoms and injected into the B plane where virtually the whole excess charge is confined, resulting in Coulomb attraction along the c-axis, i.e. between the Mg and the B layer. The charge in the boron plane is piled up in-between the boron atoms, making the boron planes, which are believed to be responsible for the superconductivity, covalent. It is interesting to note that there is also a depletion of charge out to a distance of some 0.05 nm away from the B atoms (similar charge deficiency was also observed for Mg atoms as indicated in dark-blue in fig.4). Thus electrons from the Mg as well as the B atoms participate in forming the covalent bonding in the boron plane. Qualitatively, our results agree with other experimental and theoretical studies, but comparisons of the subtle details are not feasible because structure factor values have not been published, and we are aware of only one work that presents difference maps, the x-ray study on single crystals that were reported 4.5 % low in Mg [13]. Total electron density maps are less informative, and probably less reliable without having determined the structure factors extremely far out in reciprocal space due to the truncation errors. Comparison of total electron density maps from Fourier transform based on our structure factors from DFT and the procrystal by using structure factors reaching out to the interplanar spacing of 0.05 nm, showed that the fine details were very similar, signaling that the truncation error, rather than the electron density, is responsible for these fine details. The truncation problem is virtually absent in the difference map in that there is typically only 2 % difference between the structure factors calculated based on DFT and the procrystal model. There are also similarities between our electron density map and previous x-ray studies [13], in particular the excess charge in between the B atoms, mainly associated with the $\sigma$ bonding electrons that have radial symmetry around the line joining neighboring B atoms. A small deviation from radial symmetry, attributed to the $\pi$ non-bonding electrons, is perceived by comparing fig. 3a and fig.3c.

To summarize, we have studied the valence electron distribution in the superconductor $MgB_2$ by accurately measuring structure factors that are highly sensitive to the rearrangement of charge using electron and synchrotron x-ray powder diffraction techniques and first principle calculations. Our coordinated experimental and theoretical study show that each Mg atom has donated two electrons to the boron layer, suggesting that the boron layer, in addition to having the same honeycomb structure as the carbon layers in graphite, also have the same number of valence electrons, and these electrons are mainly located in the $p_xp_y$ orbitals between neighboring boron atoms.

**ACKNOWLEDGEMENT**

This work was supported by Division of Materials Sciences, Office of Basic Energy Science, U.S. Department of Energy, under contract No. DE-AC02-98CH10886.

Table 1  Experimental structure factors from electron diffraction and compared with those calculated for overlapping atomic densities (procrystal) at room temperature.

| Reflection | | Electron structure factor | Converted to X-ray structure factor |
|---|---|---|---|
| 001 | Experiment | -0.20 | 2.15 |
|  | Procrystal | -0.91 | 2.75 |
| 100 | Experiment | 0.94 | 5.53 |
|  | Procrystal | 0.63 | 5.98 |
| 101 | Experiment | 2.60 | 10.64 |
|  | Procrystal | 2.69 | 10.42 |
| 002 | Experiment | 2.93 | 11.43 |
|  | Procrystal | 2.94 | 11.40 |

Table 2  X-ray structure factors and also those converted from experimental electron diffraction compared with calculations at zero K

| Reflection | Observations | | Calculations | |
|---|---|---|---|---|
|  | X-ray diffraction | Electron diffraction | DFT | Procrystal |
| 001 | 2.15 | 2.17 ± 0.03 | 2.19 | 2.77 |
| 100 | 5.68 | 5.61 ± 0.08 | 5.78 | 6.07 |
| 101 | 11.19 | 10.88 ± 0.15 | 10.68 | 10.65 |
| 002 | 11.24 | 11.80 ± 0.15 | 11.79 | 11.77 |

Figure captions

Fig.1

(a-c) Conventional convergent beam diffraction patterns, (a) experimental (b) calculated for procrystal and (c) best fit to the experiment. (d-f) PARODI patterns, (d) experimental (e) calculated for procrystal and (f) best fit to the experiment. Sketches of experimental setup for both methods using a wedge sample are included at the top of the figures.

Fig.2

Fitting results of Rietveld analysis of $MgB_2$ x-ray powder diffraction. (a) experimental observation, (b) difference spectrum refined as $MgB_2$ with $R_{wp}$=11.2% and $R_I$=11.5%, where

$$R_{wp} = \sqrt{\frac{\Sigma w(I'_{obs} - I'_{cal})^2}{\Sigma wI'^2_{obs}}}, \quad and \quad R_I = \frac{\Sigma(I_{obs} - I_{cal})}{\Sigma I_{obs}}.$$

(c) difference spectrum refined as MgBC with $R_{wp}$=9.7% and $R_I$=8.3%. Insets are the enlarged box-area (001, 100 and 101 reflections) in (b-c).

Fig.3

DFT calculated deformation charge density maps $\Delta\rho(\vec{r})$ with contours shown as dashed black-line for $\Delta\rho<0$ with intervals of 0.2 e/Å$^3$, orange-line for $\Delta\rho=0$, and solid black-line for $\Delta\rho>0$ with intervals of 0.05 e/Å$^3$. A color scheme of black-blue-green for below zero level, yellow-purple for at and above zero level was used. a) and b) are planes normal to the c axis through the Mg and B atoms, respectively. (c) The 110 plane through the Mg and B atoms. (d) Differs from (c) in that we have replaced the 001, 100,101 and 002 DFT calculated structure factors with these four determined by electron diffraction.

Fig.4

A three-dimensional valence electron distribution in $MgB_2$ based on precision electron diffraction measurements and DFT calculations. The map was made with the same color scheme used in fig.3. A translucency factor for $\Delta\rho$ around zero (0.1~0.1) was used to improve the visibility.

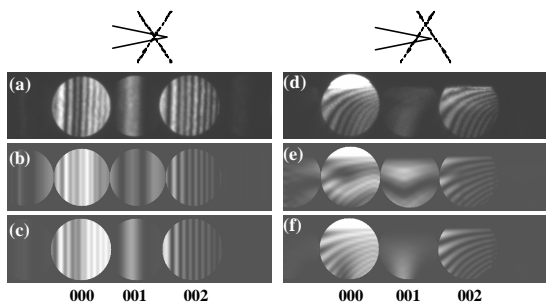

Fig.1

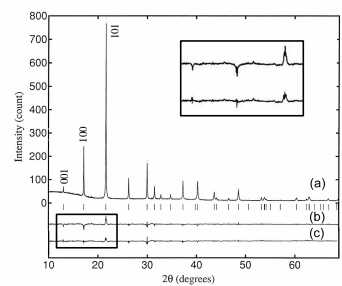

Fig.2

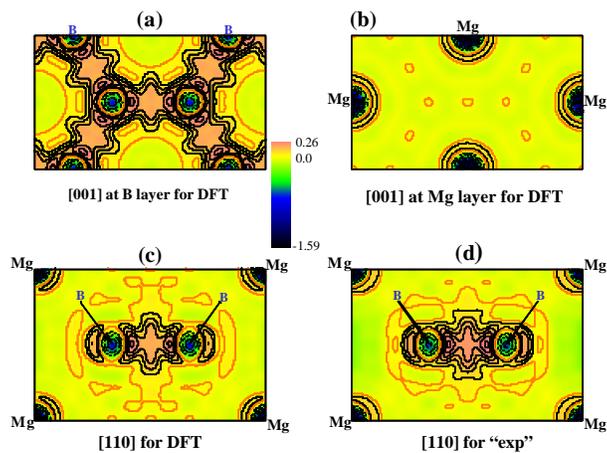

Fig.3

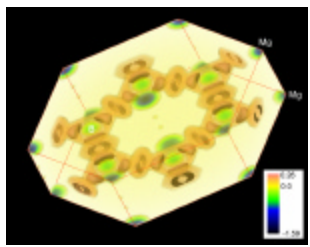

Fig.4